\documentclass[article,twocolumn]{revtex4}

\usepackage{graphicx}
\usepackage{color}

\newcommand{\ket}[1]{\left| #1 \right\rangle}
\newcommand{\bra}[1]{\left\langle #1 \right|}

\newcommand{\proj}[1]{| #1\rangle\!\langle #1 |}
\newcommand{\Tr}{\mathrm{Tr}}

\newcommand{\be}{\begin{equation}}
\newcommand{\ee}{\end{equation}}
\newcommand{\bea}{\begin{eqnarray}}
\newcommand{\eea}{\end{eqnarray}}

\newcommand{\PI}{{\rm PI}}
\newcommand{\rhopi}{\rho^{\PI}}

\begin{document}
\title{On the permutationally invariant part of a density matrix and nonseparability of $N$-qubit states}
\author{Ting Gao$^{1,3}$, Fengli Yan$^{2,3}$, S.J. van Enk$^3$}
\affiliation{
$^1$College of Mathematics and Information Science, Hebei
Normal University, Shijiazhuang 050024, China\\
 $^2$College of Physics Science and Information Engineering, Hebei Normal University, Shijiazhuang 050024,  China \\
$^3$Department of Physics, University of Oregon, Eugene, OR 97403, USA}
\begin{abstract}
We consider the concept of ``the permutationally invariant (PI) part of a density matrix,'' which has proven very useful for both efficient quantum state estimation and entanglement characterization of $N$-qubit systems. We show here that the concept is, in fact, basis-dependent, but that this basis dependence makes it an even more powerful concept than has been appreciated so far. By considering the PI part $\rhopi$ of a general (mixed) $N$-qubit state $\rho$, we obtain: (i) strong bounds on quantitative nonseparability measures,
(ii)  a whole hierarchy of multi-partite separability criteria (one of which entails a sufficient criterion for genuine $N$-partite entanglement) that can be experimentally determined by just $2N+1$ measurement settings, (iii) a definition of an efficiently measurable degree of separability, which  can be used for quantifying a novel aspect of decoherence of $N$ qubits, and (iv) an explicit example that shows there are, for increasing $N$, genuinely $N$-partite entangled states lying closer and closer to the maximally mixed state. 
Moreover, we show that if the PI part of a state is $k$-nonseparable, then so is the actual state. 
We further argue to add as requirement on any multi-partite entanglement measure $E$ that it satisfy
$E(\rho)\geq E(\rhopi)$, even though the operation that maps $\rho\rightarrow\rhopi$ is not local.
\end{abstract}
\maketitle
Quantum computing experiments are still firmly in the testing phase.
One of the practical questions that emerges in the quest to build a quantum computer
is what one should measure or verify in an experiment with a number $N$ of qubits that is too small to run, say, an unprecompiled \cite{smolin} algorithm in a fault tolerant way, but too large to be amenable to a full simulation or analysis.

Instead of performing quantum-state tomography 
on the full quantum state $\rho$ of $N$ qubits,
a useful shortcut is to perform permutationally invariant (PI) tomography \cite{toth2010,moroder2012} to estimate just the PI part of $\rho$, defined as
\be
\rhopi=\frac{1}{N!}\sum_{n=1}^{N!}
\Pi_n\rho\Pi_n^\dagger,
\ee
where the $\{\Pi_n\}$ denotes the set of all $N!$ permutations of the $N$ qubits.

There are various points to limiting oneself to this procedure:   (i) many experiments in the testing phase have as their goal the generation of PI states, with the landmark experiments
on 14 entangled ions in an ion trap (GHZ state) \cite{monz2011} and tomography on 8 ions (W state) \cite{haffner2005}, respectively, being two relevant examples, (ii) there is a simple test to verify whether one's actual state is close to a PI state \cite{toth2010},  (iii) the number of parameters 
needed to describe a PI state is ${\cal O}(N^3)$ (and so a full numerical analysis of such an experiment is feasible for at least $N=25$ \cite{schwarz2013}), (iv) the number of measurement settings needed is merely ${\cal O}(N^2)$, and so the experiment itself is quite feasible, (v) $\rhopi$ contains useful information about $\rho$: e.g., if $\rhopi$ is not fully separable (i.e., entangled), then so is $\rho$, and (vi) ground states of interesting model Hamiltonians, such as certain Hubbard models, are PI. Moreover, a lot is known about multi-partite entanglement properties of PI states and equivalence classes under SLOCC for such symmetric states \cite{lyons2011,ribeiro2011,markham2011,novo2013}, and both entanglement witnesses and criteria \cite{toth2010b} as well as tomography \cite{klimov2013} can be optimized for PI states.

We will extend this list of useful properties of the PI part of a state, but only after pointing out that the procedure of taking the PI part is basis dependent, or, perhaps more accurately, isomorphism dependent. For, the procedure of taking the PI part of a state relies on the simple fact that any two two-dimensional Hilbert spaces are isomorphic. The isomorphism is made explicit when we declare that two particular basis states of qubit $A$, say, the states $\ket{0}_A$ and $\ket{1}_A$ are physically the same as the states
$\ket{0}_B$ and $\ket{1}_B$ of qubit $B$, respectively 
\footnote{{\em Permutation} symmetry must be distinguished from {\em exchange} symmetry, which plays a fundamental role in quantum statistics. In the latter case one considers the effect on a wave function of the exchange of {\em un}physical labels of identical particles. Here, in contrast, we distinguish the particles by their locations, and the labels $A$ and $B$ refer to spatially separated systems and are, therefore, physical.}. This is such an obvious convention that it is always left unstated. 

That it nonetheless has some nontrivial consequences, is shown by the following example. The state $\ket{\psi}:=\ket{01}+i\ket{10}$ is maximally entangled, but the PI part of this state is
an equal mixture of $\ket{\psi}$ and
$\ket{\phi}:=\ket{10}+i\ket{01}$, and that mixture is not entangled. However, had we made a slight change in the choice of basis states for the first qubit by
picking $\{\ket{0},i\ket{1}\}$, then the PI part of $\ket{\psi}$ would have been maximally entangled.
In fact, for any {\em pure} 2-qubit state, adjusting the basis states so as to be aligned with the Schmidt basis shows that its PI part is always just as much entangled as the original state itself.

We can generalize this example to obtain a much stronger statement. Since the PI part of a state is obtained by mixing states that all have exactly the same (multi-partite) entanglement properties, any quantitative measure of (multi-partite) entanglement {\em ought to} satisfy
\be
E(\rho)\geq E(\rhopi).
\ee
This bound may be quite weak: the example above shows that the right-hand side (rhs) may be zero, even when $\rho$ is maximally entangled.
However, by maximizing the rhs over all possible single-qubit basis changes $B:=\otimes_{k=1}^NU_k$, we obtain the much stronger bound
\be
E(\rho)\geq \max\limits_{\texttt{all bases}\, B}  E(\rho^{(\PI)_B}).
\ee
While each entanglement measure ought to satisfy these relations, one still has to prove for any individual proposed measure that it indeed does. Note that an interesting and useful intermediate symmetry  [intermediate between PI with the basis fixed and PI with arbitrary local bases] for $N$-qubit states was considered in Refs.~\cite{eltschka2012,siewert2012}: the so-called GHZ symmetry. This in turn leads to a similar bound on entanglement in terms of entanglement of GHZ-symmetric states \cite{eltschka2012b}.

As an illustration we will now focus on a particular class of separability criteria and measures (which, as it turns out, can be very efficiently measured). Multi-partite nonseparability is to be distinguished from multi-partite entanglement, as follows (see \cite{huber2010}). A pure $N$-partite state $\proj{\psi}$ is called $k$-separable if the $N$ parties can be partitioned into $k$ groups $A_1, A_2,\ldots, A_k$ such that the state can be written as a tensor product
$\rho_{A_1}\otimes\rho_{A_2}\ldots\otimes\rho_{A_k}$. A general mixed state $\rho$ is $k$-separable if it can be written as a mixture of $k$-separable states, and a state is $k$-nonseparable if it is not $k$-separable.
One relation between entanglement
and $k$-nonseparability is, that 2-nonseparability is equivalent to genuine $N$-partite entanglement. 

A quantitative measure of multi-partite $k$-inseparability (with computable lower bounds) was introduced in \cite{hong2012} (which generalizes to arbitrary $k$ the measure for $k=2$ introduced in \cite{ma2011}). For a pure $N$-partite state $\ket{\psi}$, one may define the so-called $k$-ME concurrence 
in terms of all possible $k$-partitions, as follows:
\be
C_{k-\mathrm{ME}}(\proj{\psi})=
\min\limits_A\sqrt{\frac{2\sum_{t=1}^k[1-\Tr(\rho_{A_t}^2)]}{k}},
\ee 
where $\rho_{A_t}=\Tr_{\bar{A_t}}(\proj{\psi})$ is the reduced density matrix of subsystem $A_t$ (and $\bar{A}_t$ is the complement of $A_t$). One then uses the standard convex-roof construction to extend the definition of $k$-ME concurrence
to mixed states.  One important property is
this: $C_{k-\mathrm{ME}}(\proj{\psi})$ is nonzero if and only if $\ket{\psi}$ is $k$-nonseparable (and so it equals zero if and only if $\ket{\psi}$ is $k$-separable).

We state now:

\textbf{Theorem 1} ~~ Suppose that $\rho$ is an $N$-partite state. Then the maximum of $k$-ME concurrence \cite{hong2012} of the permutational invariant part of $\rho$ is a lower bound on the $k$-ME concurrence \cite{hong2012} of the original state $\rho$. That is:
\begin{equation}\label{}
C_{k-\mathrm{ME}}(\rho)\geq\max\limits_{\texttt{all bases}\,B} C_{k-\mathrm{ME}}(\rho^{(\PI)_B}).
\end{equation}
The proof is given in the Supplementary Material. An immediate consequence (which can be proven in other ways, too) of this Theorem is:

\textbf{Corollary} ~~ If  an $N$-partite state $\rho$ is $k$-separable,  then so is its PI part. Conversely, if the permutationally invariant part $\rhopi$ of an $N$-partite state $\rho$ is $k$-nonseparable, then so is the actual state $\rho$.

(Note that for $k=N$ this result was known (it is stated in Ref.~\cite{toth2010}).)

Next, we focus on separability criteria that work very well for, e.g., noisy versions of both GHZ states and W states \cite{gao2010,gao2011,gao2012k} and that allow for a simple and inexpensive test for the whole hierarchy of $k$-separability with $k$ running from $N$ down to 2. 
If an $N$-qubit density matrix $\rho$ is $k$-separable, then its matrix elements (in any given orthonormal basis) have been shown to fulfill the condition \cite{gao2012k}
\bea\label{THE}
A\leq B+C\frac {N-k}{2},
\eea
where we defined 
\bea
A&:=&\sum\limits_{0\leq i< j\leq N-1}|\rho_{2^i+1, 2^j+1}|,\nonumber\\
B&:=&\sum\limits_{0\leq i< j\leq N-1}\sqrt{\rho_{1,1}\rho_{2^i+2^j+1,2^i+2^j+1}},\nonumber\\
C&:=&\sum\limits_{i=0}^{N-1}\rho_{2^i+1,2^i+1}.
\eea
If the above inequality is violated, then $\rho$ is necessarily $k$-nonseparable (i.e., not $k$-separable). In particular, if the inequality for $k=2$ is violated, the state must be genuinely $N$-partite entangled. (We repeat that the notion of $k$-nonseparability is distinct from $k'$-partite entanglement, except for the case $k=2$ and $k'=N$.)
As was shown in \cite{gao2012k}, the number of local measurement settings needed to determine all matrix elements appearing in the inequality (\ref{THE})
is just $(5N^2-3N+2)/2$.
The number of parameters needed to fully describe such an experiment on a generic $N$-qubit state would still grow exponentially, as ${\cal O}(2^N N^2)$. 

By restricting ourselves to PI states (or, in fact, to the PI part of the full density matrix) we can reduce both the number of measurement settings needed and, by much more, the number of parameters 
needed to describe the experiment.
To see this, first write a general PI state of $N$ qubits as
\be\label{Yan}
\rhopi=\sum\limits_{k,l,m,n\geq 0 \atop k+l+m+n=N} e_{klmn}\Pi(\sigma_x^{\otimes k}\otimes \sigma_y^{\otimes l}\otimes \sigma_z^{\otimes m}
\otimes \openone^{\otimes n}),
\ee
where $\Pi=\sum_n\Pi_n$ represents the sum of all $N!$ permutations on $N$ qubits.
We now rewrite the matrix elements needed for criterion (\ref{THE}) in terms of the coefficients $e_{klmn}$.
The off-diagonal elements appearing on the left-hand side of  (\ref{THE}) can be written (after some straightforward, even if a little bit tedious, algebra) as 
\begin{equation}
\rhopi_{2^i+1, 2^j+1}|_{i\neq j}
=2(N-2)!\sum _{m=0 }^{N-2}(e_{20m,N-2-m}+e_{02m,N-2-m}).
\end{equation}
The values of these matrix elements, therefore, do not actually depend on the indices $i,j$, and they can be determined from
just $e_{20mn}$ and $e_{02mn}$ for $m+n=N-2$.
One obtains similarly the diagonal elements appearing on the right-hand side of (\ref{THE}) as
\be\label{11}
\rhopi_{1,1}=N!\sum_{m=0}^N e_{00m,N-m},
\ee
as well as
\be\label{2i}
\rhopi_{2^i+1,2^i+1}=
(N-1)!\sum_{m=0}^N (N-2m)e_{00m,N-m}.
\ee
Finally, for $i\neq j$ we have
\bea\label{2ij}
\rhopi_{2^i+2^j+1,2^i+2^j+1}=\nonumber\\
(N-2)!\sum\limits_{m=0}^{N}[(N-2m)^2-N] e_{00m,N-m}.
\eea
All of this tells us which of the coefficients $e_{klmn}$ we have to determine (measure) in order to be able to apply the $k$-separability criterion to a PI state. Noting that $e_{000N}$ is fixed by normalization, we only need the following three types of coefficients
\begin{enumerate}
\item
 $e_{00m,N-m}$ for $m=1,2,\ldots, N$,
 \item $e_{20m,N-m-2}$ for $m=0,1,\ldots, N-2$,
 \item $e_{02m,N-m-2}$ for $m=0,1,\ldots, N-2$,
 \end{enumerate}
 and there are $3N-2$ such coefficients.
We wish to determine all of these by measurements of local observables of the form $\hat{A}^{\otimes N}$.
It was shown in Ref.~\cite{toth2010} that $(N+2)(N+1)/2$ such measurements suffice to determine the PI part of any $N$-qubit state, and thereby {\em all} coefficients $e_{klmn}$.
One expects that for our more modest purpose fewer measurements are required.

We first note that measuring $\sigma_z$ on all qubits provides us with the first type of coefficients, since
 \begin{equation}
\Tr\rhopi\Pi(\sigma_z^{\otimes m}\otimes I^{\otimes (N-m)})= e_{00m,N-m} {(N-m)!}2^N N!m!.
\end{equation}
For the remaining coefficients, consider the measurement of the observable $\hat{A}^{\otimes N}$ with
$\hat{A}=a\sigma_x+b\sigma_y+c\sigma_z$ and $a^2+b^2+c^2=1$.
We note that
\bea
\Tr\rhopi \Pi( \hat{A}^{\otimes (N-n)}\otimes \openone^{\otimes n})=\nonumber\\
\sum_{k,l,m\atop{k+l+m=N-n}}e_{klmn}
(N-n)!a^kb^lc^m2^NN!n!.
\eea
Specializing to the case $a=0$, we get
\bea
&&\Tr\rhopi\Pi(\hat{A}_{a=0}^{\otimes (N-n)}\otimes \openone^{\otimes n})= \nonumber\\
&&e_{0,0,N-n,n} {(N-n)!c^{N-n}}2^N N!n!\nonumber\\
&+&e_{0,1,N-n-1,n} {(N-n)!bc^{N-n-1}}2^N N!n!\nonumber\\
&+&e_{0,2,N-n-2,n} {(N-n)!b^{2}c^{N-n-2}}2^N N!n!\nonumber\\
&+&\sum\limits_{l=3 }^{N-n} e_{0,l,N-n-l,n} {(N-n)!b^{l}c^{N-n-l}}2^N N!n!.
\eea
The coefficients appearing on the first line of the right-hand side in this equation are already determined from the measurement of $\sigma_z^{\otimes N}$.
There remain then $N-n$ linear equations for $N-n$ unknowns, for each $n=0,1,\ldots, N-2$ (for $n=N-1$ and $n=N$ no useful information is obtained). It suffices, therefore, to measure $N$ different observables with $a=0$ to determine the coefficients $e_{02,N-n-2,n}$ for each $n$, by linear inversion. 
Similarly, it suffices to measure $N$ other observables with $b=0$ to determine $e_{20,N-n-2,n}$.
Hence in total $2N+1$ measurement settings suffice to determine everything we need for all $k$-separability criteria (\ref{THE}). 
(We show explicitly how one may implement this procedure for $N=3$ in the Supplementary Material.)

The number of coefficients $e_{klmn}$ of a general PI state that such an experiment with $2N+1$ measurement settings actually determines (including many that are not needed to evaluate (\ref{THE}))
is 
\be
N+2\sum_{n=0}^{N-1}(N-n)=N^2+2N.
\ee
This, then, equals the number of model parameters one needs to describe this experiment in terms of PI states.
Note that, since the number of independent PI measurement outcomes
for one measurement setting  equals $N$ (the number of times one finds ``spin up''), not all measurement outcomes corresponding to the $2N+1$ measurement settings are independent.

More ambitiously, one may measure all ${\cal O}(N^3)$ coefficients that determine a general PI state by running ${\cal O}(N)$ experiments of the above type (with $2N+1$ measurement settings), where each 
experiment determines all the coefficients needed for the $k$-separability criteria {\em in a particular basis}.

By inverting the relation (\ref{THE}) we can define an effective $k_{{\rm eff}}$ as follows
\bea
k_{{\rm eff}}:=N-2\frac{A-B}{C}.
\eea
This number is efficiently measurable and quantifies the degree of separability of the PI part of one's actual state. For example, whenever one measures $k_{{\rm eff}}<2$ the underlying PI quantum state must be genuinely $N$-partite entangled. For a PI mixture of fully separable $N$-qubit states one should find that  $k_{{\rm eff}}\geq N$.

One can now envisage an experiment in which the degree of separability is varied and measured. For example, along the lines of
 the experiment of Ref.~\cite{choi2010}, one may generate a genuinely $N$-partite entangled PI state, and then observe its decay over time from one separability class to the next, with $k_{{\rm eff}}$ monotonically increasing with the amount of decoherence. Or, likewise along the lines of Ref.~\cite{choi2010}, one may introduce certain imperfections on purpose, in a controlled way, so that one knows that the state one generates falls at best into
a certain class of $k$-inseparability. One ought to measure then that $k_{{\rm eff}}>k$. 

Moreover, by running the above-mentioned ${\cal O}(N)$ versions of the same experiment, determining 
the $k$-separability criteria in  different bases, one can define just as many
different degrees of separability, each one of which ought to increase over time, and each one of which has to obey the inequality $k_{{\rm eff}}>k$ if one creates the right type of imperfections.
This way one verifies directly both the theory and one's experimental control, all at a cost growing only polynomially with $N$.

\begin{figure}[t]
\includegraphics[width=2.5in]{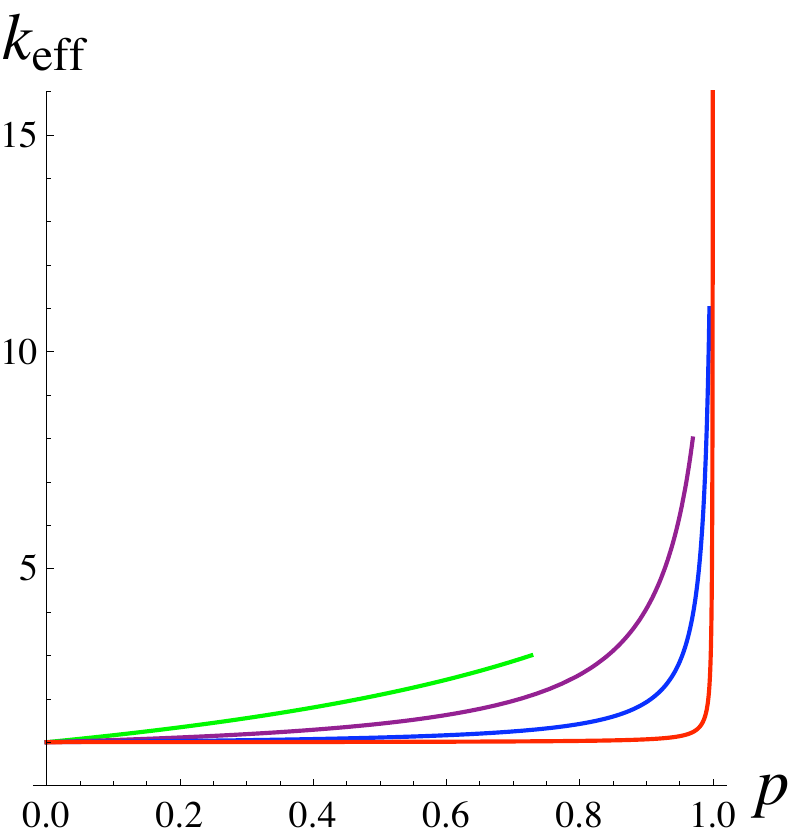} 
\caption{(Color
online). Illustration of the
degree of separability $k_{\mathrm{eff}}$. It  is plotted for the state
$\rho_N(p)$ as defined in Eq.(\ref{rhoNp}) as a function of $p$, for $N=3,8,11,16$ (corresponding to the green, purple,
blue,  and red lines, respectively; or, for color blind readers, corresponding to the top to bottom curves, respectively.). For any $N$, a value of $k_{{\rm eff}}>N$ is meaningless, and in the plot such values are left out.}
\label{plot1}
\end{figure}

Let us consider here a simple example of how $k_{{\rm eff}}$ increases with the amount of decoherence, and how it behaves as a function of the number of qubits. Consider a mixture of the $N$-qubit W state and the maximally mixed $N$-qubit state,
\be\label{rhoNp}
\rho_N(p):=(1-p)\ket{W}_N\bra{W}+p\openone/2^N,
\ee
for $0\leq p\leq 1$. For this state we obtain the degree of separability as a function of $p$ and $N$ as
\begin{equation}\label{keffpN}
k_{\mathrm{eff}}(p,N)=\frac {2^N -\left(2^N + N - 2 N^2 \right) p} {2^N + \left(-2^N + N \right) p}.
\end{equation}
We plot $k_{\mathrm{eff}}(p,N)$ as a function of $p$ in Fig.~\ref{plot1}, for several values of $N$.

One notable feature visible in the plot is that for large $N$ (even for $N\geq16$) the state $\rho_N(p)$ stays genuinely $N$-partite entangled even when $p$ is very close to unity (because $k_{\mathrm{eff}}(p,N)\approx 1$), but then $k_{\mathrm{eff}}$ suddenly shoots up to its maximum value for $p\rightarrow 1$. 
That is, in the limit of $N\rightarrow\infty$ one can find arbitrarily close to the maximally mixed (and fully separable) state a genuinely $N$-partite entangled state.
This peculiar behavior agrees with one particular fact about bi-partite entanglement in infinite dimensional systems: arbitrarily close to any separable state lies an entangled state with an arbitrarily large amount of entanglement \cite{eisert2002,enk2003}.

In conclusion, we obtained strong bounds on a multi-partite nonseparability measure, by considering the basis dependence of the permutation operation. 
We showed that whenever the permutationally invariant (PI) part of a state is $k$-nonseparable, then so is the state itself. Moreover, we have shown it takes just $2N+1$ measurement settings to determine all matrix elements needed for checking all $k$-separability criteria for $k=2,3,\ldots, N$ obtained in \cite{gao2012k}, as applied to PI $N$-qubit states. The required measurements can be performed {\em and} analyzed (using, e.g., model selection \cite{burnham2002}) for a number of qubits 
that goes well beyond the current state-of-the-art of $N=14$.
From these $k$-separability criteria we obtained in a natural way ``degrees of separability,'' which are efficiently measurable even for many qubits, and which ought to increase monotonically with the amount of decoherence.

\section*{Acknowledgements}
This work was supported by the National Natural Science Foundation
of China under Grant No: 11371005, and by the Hebei Natural Science Foundation
of China under Grant No: A2012205013.
\bibliography{perm_ent_4}

\end{document}


\section*{Supplementary Material}
We first  consider the proof of Theorem 1.
\subsection*{Proof of Theorem 1}
Note that for any $k$-partition $A_1|A_2|\cdots|A_k$ (of $\{1,2,\cdots, n\}$),  $\Pi_n(A_1)|\Pi_n(A_2)|\cdots|\Pi_n(A_k)$ is also a $k$-partition (of $\{1,2,\cdots, N\}$), and for any $N$-partite pure state $|\psi\rangle$, $\Pi|\psi\rangle$ is also a pure state. The reduced density matrix $\sigma_{A_i}$ is equal to the reduced density matrix $\rho_{\Pi_n(A_i)}$, i.e., $\sigma_{A_i}=\rho_{\Pi_n(A_i)}$,  which implies that
\begin{equation}\label{}
C_{k-\mathrm{ME}}(\Pi_n|\psi\rangle)=C_{k-\mathrm{ME}}(|\psi\rangle),
\end{equation}
where $\rho=|\psi\rangle\langle\psi|$ and $\sigma=\Pi_n|\psi\rangle\langle\psi|\Pi_n^\dagger$. Here $\Pi_n$ is an arbitrary permutation of $N$ particles $\{1,2,\cdots, N\}$. The permutationally invariant (PI) part of $\rho=|\psi\rangle\langle\psi|$ is
\begin{equation}\label{}
\rhopi=\big(|\psi\rangle \langle\psi|\big)^{\PI}=\sum_{\Pi_n\in S_N}\frac{1}{N!}\Pi_n|\psi\rangle \langle\psi|\Pi_n ^\dagger.
\end{equation}
Here $S_N=\{\Pi_n: ~ \Pi_n ~ \textrm{is permutation of} ~ N ~ \textrm{particles}\}$ is the symmetric group of all permutations on $N$ particles.
By the definition of $k$-ME concurrence
we have
\bea
C_{k-\mathrm{ME}}(\rhopi)\leq\sum_{\Pi_n\in S_N}\frac{1}{N!}C_{k-\mathrm{ME}}(\Pi_n|\psi\rangle)=\nonumber\\
\sum_{\Pi_n\in S_N}\frac{1}{N!}C_{k-\mathrm{ME}}(|\psi\rangle)=C_{k-\mathrm{ME}}(|\psi\rangle).
\eea
It
 follows that for any pure state $\rho=|\psi\rangle\langle\psi|$ 
\begin{equation}\label{}
C_{k-\mathrm{ME}}(\rhopi)\leq C_{k-\mathrm{ME}}(\rho).
\end{equation}
For any $N$-partite mixed state $\rho=\sum\limits_{m}p_m|\psi_m\rangle \langle\psi_m|$, we obtain
\bea
\rhopi&=&\frac{1}{N!}\sum_{\Pi_n\in S_N}\Pi_n\rho\Pi_n ^\dagger =\nonumber\\
&&\sum\limits_{m}p_m\Big(\sum_{\Pi_n\in S_N}\frac{1}{N!}\Pi_n|\psi_m\rangle \langle\psi_m|\Pi_n^\dagger\Big)=\nonumber\\
&&\sum\limits_{m}p_m\big(|\psi_m\rangle \langle\psi_m|\big)^\PI,
\eea
and from the convexity of $k$-ME concurrence we have then
\bea
C_{k-\mathrm{ME}}(\rhopi)\leq \sum\limits_{m}p_mC_{k-\mathrm{ME}}\Big(\big(|\psi_m\rangle \langle\psi_m|\big)^\PI\Big)
\nonumber\\
\leq  \sum\limits_{m}p_mC_{k-\mathrm{ME}}\big(|\psi_m\rangle\big).
\eea
This implies that for mixed states, too, we have
\begin{equation}\label{}
C_{k-\mathrm{ME}}(\rhopi)\leq C_{k-\mathrm{ME}}(\rho).
\end{equation}
Since the above proof is valid for any choice of (local)  bases, we obtain Theorem 1:
\begin{equation}\label{}
\max\limits_{\texttt{all bases}\,\,B} C_{k-\mathrm{ME}}(\rho^{(\PI)_B})\leq C_{k-\mathrm{ME}}(\rho),
\end{equation}
where $B=U_1\otimes U_2\otimes\ldots U_N$ denotes an arbitrary local basis change for each party.
\subsection*{Choosing seven observables for three qubits}
In the main text we showed that $2N+1$ appropriate measurement settings can be found and that linear equations can be inverted so as to
reconstruct all those density matrix elements that are needed for the evaluation of all our $k$-separability criteria for PI $N$-qubit states. Here we show explicitly how this works for $N=3$. In fact, we show here that (amongst many other possible choices) the following 7  observables of the form $\hat{A}_t^{\otimes 3}$ for $t=1,2,\ldots, 7$ suffice for 3 qubits:
\bea
&&\hat{A}_1=\sigma_z;\,\,\,
\hat{A}_2=\sigma_x;\,\,\,
\hat{A}_3=\sigma_y;\nonumber\\
&&\hat{A}_{4,5}=\frac{\sigma_y\pm\sigma_z}{\sqrt{2}};\,\,
\hat{A}_{6,7}=\frac{\sigma_x\pm\sigma_z}{\sqrt{2}}.
\eea
In order to demonstrate this, we start by noting that a general permutationally invariant density matrix $\rho$ of 3 qubits can be written in the form
\begin{widetext}
\begin{equation}\label{observable1}
\begin{array}{ll}
\rho= & \frac {1}{48}\Pi(I\otimes I\otimes I)+ e_{1002}\Pi(\sigma_x\otimes I\otimes I)+e_{0102}\Pi(I\otimes \sigma_y\otimes I)+e_{0012}\Pi(I\otimes I\otimes \sigma_z)\\
& +e_{2001}\Pi(\sigma_x\otimes \sigma_x\otimes I)
+e_{0201}\Pi(\sigma_y\otimes \sigma_y\otimes I)
+e_{0021}\Pi(\sigma_z\otimes \sigma_z\otimes I)
+e_{1101}\Pi(\sigma_x\otimes \sigma_y\otimes I) \\
& +e_{1011}\Pi(\sigma_x\otimes \sigma_z\otimes I)
+e_{0111}\Pi(\sigma_y\otimes \sigma_z\otimes I)
+e_{3000}\Pi(\sigma_x\otimes \sigma_x\otimes \sigma_x)
+e_{2100}\Pi(\sigma_x\otimes \sigma_x\otimes \sigma_y)\\
& +e_{2010}\Pi(\sigma_x\otimes \sigma_x\otimes \sigma_z)
+e_{1110}\Pi(\sigma_x\otimes \sigma_y\otimes \sigma_z)
+e_{1200}\Pi(\sigma_x\otimes \sigma_y\otimes \sigma_y)
+e_{1020}\Pi(\sigma_x\otimes \sigma_z\otimes \sigma_z)\\
& +e_{0300}\Pi(\sigma_y\otimes \sigma_y\otimes \sigma_y)
+e_{0210}\Pi(\sigma_y\otimes \sigma_y\otimes \sigma_z)
+e_{0120}\Pi(\sigma_y\otimes \sigma_z\otimes \sigma_z)
+e_{0030}\Pi(\sigma_z\otimes \sigma_z\otimes \sigma_z).
\end{array}
\end{equation}
\end{widetext}
Here $\Pi$ denotes the summation of all 6 permutations of 3 qubits. For example, $\Pi(\sigma_x\otimes \sigma_y\otimes \sigma_z)=\sigma_x\otimes \sigma_y\otimes \sigma_z+\sigma_x\otimes \sigma_z\otimes \sigma_y+\sigma_y\otimes \sigma_z\otimes \sigma_x+\sigma_y\otimes \sigma_x\otimes \sigma_z+ \sigma_z\otimes \sigma_x\otimes \sigma_y+\sigma_z\otimes \sigma_y\otimes \sigma_x$.
In the computational basis, we have
\begin{equation}
\rho_{1,1}= \langle 000|\rho|000\rangle=\frac {1}{8}+6e_{0012}+6e_{0021}+6e_{0030},
\end{equation}
as well as
\begin{equation}\begin{array}{ll}
                   & \rho_{2^0+1,2^0+1}=\langle 001|\rho|001\rangle =\\
                & \rho_{2^1+1,2^1+1}=\langle 010|\rho|010\rangle =\\
                 & \rho_{2^2+1,2^2+1}=\langle 100|\rho|100\rangle =\\
                  & \frac {1}{8}+2e_{0012}-2e_{0021}-6e_{0030},
                \end{array}
\end{equation}
and
\begin{equation}
\begin{array}{ll}
  & \rho_{2^1+2^0+1,2^1+2^0+1}=\langle 011|\rho|011\rangle= \\
 & \rho_{2^2+2^1+1,2^2+2^1+1}=\langle 110|\rho|110\rangle =\\
 & \rho_{2^2+2^0+1,2^2+2^0+1}=\langle 101|\rho|101\rangle =\\
 & \frac {1}{8}-2e_{0012}-2e_{0021}+6e_{0030},
\end{array}
\end{equation}
and 
\begin{equation}
\begin{array}{ll}
  & \rho_{2^0+1,2^1+1}=\langle 001|\rho|010\rangle =\\
  & \rho_{2^0+1,2^2+1}=\langle 001|\rho|100\rangle =\\
  & \rho_{2^1+1,2^2+1}=\langle 010|\rho|100\rangle =\\
  & 2e_{2001}+2e_{0201}+2e_{2010}+2e_{0210}.
\end{array}
\end{equation}
The diagonal elements in $B$ and $C$ can be determined by measuring  $\sigma_z\otimes\sigma_z\otimes\sigma_z$, since
\bea
\langle \Pi(\sigma_z\otimes\sigma_z\otimes\sigma_z)\rangle=\mathrm{Tr}\{\rho \Pi(\sigma_z\otimes\sigma_z\otimes\sigma_z\}=288e_{0030},
\eea
and 
\bea
\langle \Pi(\sigma_z\otimes\sigma_z\otimes I)\rangle=\mathrm{Tr}\{\rho \Pi(\sigma_z\otimes\sigma_z\otimes I\}=96e_{0021},
\eea
and 
\bea
\langle \Pi(\sigma_z\otimes I\otimes I)\rangle=\mathrm{Tr}\{\rho \Pi(\sigma_z\otimes I\otimes I\}=96e_{0012}.
\eea
The off-diagonal elements $e_{2001}, e_{0201}$ in $A$ can be obtained by measuring $\sigma_x^{\otimes 3}, \sigma_y^{\otimes 3}$, respectively, while the other off-diagonal elements $e_{2010}, e_{0210}$ in $A$ can be obtained by measuring $ \sigma_z^{\otimes 3}, (\frac{\sqrt{2}}{2}\sigma_y+\frac{\sqrt{2}}{2}\sigma_z)^{\otimes 3},  (\frac{\sqrt{2}}{2}\sigma_y-\frac{\sqrt{2}}{2}\sigma_z)^{\otimes 3},(\frac{\sqrt{2}}{2}\sigma_x+\frac{\sqrt{2}}{2}\sigma_z)^{\otimes 3},  (\frac{\sqrt{2}}{2}\sigma_x-\frac{\sqrt{2}}{2}\sigma_z)^{\otimes 3}$. The details are as follows:
Let
\begin{equation}
\hat{A}=a\sigma_x+b\sigma_y+c\sigma_z,\end{equation}
with $(a,b,c)$ a unit vector in 3-dimensional Euclidean space.  Then we have
\begin{widetext}
\begin{equation}
\begin{array}{ll}
 \hat{A}^{\otimes 3}&=
  \frac {1}{ 6}a^3\Pi(\sigma_x^{\otimes 3})+\frac {1}{ 6}b^3\Pi(\sigma_y^{\otimes 3})
+\frac {1}{6}c^3\Pi(\sigma_z^{\otimes 3})+\frac {1}{2}b^2c\Pi(\sigma_y^{\otimes 2}\otimes \sigma_z)
+\frac {1}{ 2}a^2b\Pi(\sigma_x^{\otimes 2}\otimes \sigma_y)\\
&+\frac {1}{ 2}a^2c\Pi(\sigma_x^{\otimes 2}\otimes \sigma_z)+\frac {1}{ 2}ab^2\Pi(\sigma_x\otimes \sigma_y^{\otimes 2})+\frac {1}{ 2}ac^2\Pi(\sigma_x\otimes \sigma_z^{\otimes 2})+\frac {1}{ 2}bc^2\Pi(\sigma_y\otimes \sigma_z^{\otimes 2})+abc\Pi(\sigma_x\otimes \sigma_y\otimes \sigma_z).\end{array}\end{equation}
\end{widetext}
Let $a=0$, so that
\bea
 \hat{A}_{a=0}^{\otimes 3}
&=&\frac {1}{ 6}b^3\Pi(\sigma_y^{\otimes 3})
+\frac {1}{6}c^3\Pi(\sigma_z^{\otimes 3})\nonumber\\
&&+\frac {1}{2}b^2c\Pi(\sigma_y^{\otimes 2}\otimes \sigma_z)
+\frac {1}{ 2}bc^2\Pi(\sigma_y\otimes \sigma_z^{\otimes 2}).
\eea
It follows that
\begin{equation}
\langle \hat{A}_{a=0}^{\otimes 3}\rangle
=48[b^3e_{0300}
+c^3e_{0030}
+b^2ce_{0210}
+bc^2 e_{0120}].
\end{equation}
This implies that
\begin{equation}
b^3e_{0300}
+b^2ce_{0210}
+bc^2 e_{0120}=\frac{1}{48}\langle (b\sigma_y+c\sigma_z)^{\otimes 3}\rangle-c^3e_{0030}.
\end{equation}
Note that $b,c$ are real numbers satisfying $b^2+c^2=1$.  Let $(a,b,c)$ be any three triples $(0,b_i,c_i)$ ($i=1,2,3$) with $b_i^2+c_i^2=1$ such that the  determinant 
 of the following 3 linear equations
 \begin{widetext}
\begin{equation}
\begin{array}{ll}
b_1^3e_{0300}+b_1^2c_1e_{0210}+b_1c_1^2 e_{0120}= & \frac{1}{48}\langle (b_1\sigma_y+c_1\sigma_z)^{\otimes 3}\rangle-c_1^3e_{0030}, \\
b_2^3e_{0300}+b_2^2c_2e_{0210}+b_2c_2^2 e_{0120}= & \frac{1}{48}\langle (b_2\sigma_y+c_2\sigma_z)^{\otimes 3}\rangle-c_2^3e_{0030}, \\
b_3^3e_{0300}+b_3^2c_3e_{0210}+b_3c_3^2 e_{0120}= & \frac{1}{48}\langle (b_3\sigma_y+c_3\sigma_z)^{\otimes 3}\rangle-c_3^3e_{0030}
\end{array}
\end{equation}
 \end{widetext}
is nonzero. We can then obtain nontrivial solutions $e_{0300}$, $e_{0210}$, $e_{0120}$.   For example, let $(a,b,c)=(0,1,0),(0,\frac{\sqrt{2}}{2}, \frac{\sqrt{2}}{2}), (0,\frac{\sqrt{2}}{2}, -\frac{\sqrt{2}}{2})$. Then we obtain the solution
\bea
e_{0210} =\frac{\sqrt{2}}{48}\left(\langle(\frac{\sigma_y+\sigma_z}{\sqrt{2}})^{\otimes 3}\rangle
-\langle (\frac{\sigma_y-\sigma_z}{\sqrt{2}})^{\otimes 3}\rangle-\frac{1}{\sqrt{2}}\langle\sigma_z^{\otimes 3}\rangle\right).\nonumber
\eea
 Thus, $e_{0210}$ can be determined by measuring  the three observables $(\frac{\sqrt{2}}{2}\sigma_y\pm\frac{\sqrt{2}}{2}\sigma_z)^{\otimes 3}$ and  $\sigma_z^{\otimes 3}$.
Similarly, we can obtain $e_{2010}$ by measuring $(\frac{\sqrt{2}}{2}\sigma_x\pm\frac{\sqrt{2}}{2}\sigma_z)^{\otimes 3}$ and $\sigma_z^{\otimes 3}$.